\documentclass[prb,twocolumn,groupedaddress,showpacs]{revtex4}

\usepackage[dvips]{graphicx}
\usepackage{color}
\usepackage{latexsym}
\usepackage{amsmath}
\usepackage{amssymb}

\begin{document}

\preprint{LA-UR-06-2505}

\title
   {Valence-Band UPS, 6$p$ Core-Level XPS Photoemission Spectroscopy, \\
   and Low-Energy Electron Diffraction of a Uranium (001) Single Crystal}

\author{C.P. Opeil}
\affiliation{Materials Science and Technology Division,
   Los Alamos National Laboratory,
   Los Alamos, NM 87545}
\author{R.K. Schulze}
\affiliation{Materials Science and Technology Division,
   Los Alamos National Laboratory,
   Los Alamos, NM 87545}

\author{B. Mihaila}
\affiliation{Theoretical Division,
   Los Alamos National Laboratory,
   Los Alamos, NM 87545}
\author{K.B. Blagoev}
\affiliation{Theoretical Division,
   Los Alamos National Laboratory,
   Los Alamos, NM 87545}
\author{R.C. Albers}
\affiliation{Theoretical Division,
   Los Alamos National Laboratory,
   Los Alamos, NM 87545}

\author{M.E. Manley}
\affiliation{Materials Science and Technology Division,
   Los Alamos National Laboratory,
   Los Alamos, NM 87545}

\author{J.C. Lashley}
\affiliation{Materials Science and Technology Division,
   Los Alamos National Laboratory,
   Los Alamos, NM 87545}

   \author{W.L. Hults}
   \affiliation{Materials Science and Technology Division,
   Los Alamos National Laboratory,
   Los Alamos, NM 87545}

\author{R.J. Hanrahan, Jr.}
\affiliation{Materials Science and Technology Division,
   Los Alamos National Laboratory,
   Los Alamos, NM 87545}

\author{J.L. Smith}
\affiliation{Materials Science and Technology Division,
   Los Alamos National Laboratory,
   Los Alamos, NM 87545}

\author{P.B. Littlewood}
\affiliation{Cavendish Laboratory,
   Madingley Road,
   Cambridge CB3 0HE,
   United Kingdom}

\date{\today}

\begin{abstract}
Valence-band ultraviolet photoemission spectroscopy (UPS) at 173K
and 6p core-level X-ray photoemission spectroscopy (XPS) at room
temperature were performed on a high quality uranium single
crystal. Significant agreement is found with first-principles
electronic band-structure calculations, using a generalized gradient
approximation (GGA). In addition, using Low Energy Electron
Diffraction (LEED) for the (001) surface, we find a well-ordered
orthorhombic crystallographic structure representative of the bulk
material.
\end{abstract}


\maketitle

\section{Introduction}

The actinide series of elements and their compounds~\cite{Schneider22} exhibit unusual
but similar properties related to the collective states of their
strongly correlated electrons. As one moves across this row of the
periodic table, electron-electron correlations increase until, at
Am, the 5$f$ electrons localize.  Uranium is interesting, since it
is believed to be in the normal itinerant (band-structure-like)
limit, where correlations may be slightly larger than usual, but do
not change the fundamental metallic nature of the material.
Nonetheless, there are tantalizing hints (anomalies) that
correlations are still playing an important role in this material.
For example, the specific heat enhancements are significantly large
compared with band-structure calculations (see below), and the
phonon spectra is strongly and anomalously softened at high
temperatures~\cite{manley1}. For this reason it is important to
explore the experimental electronic structure of U in detail and to
compare with band-structure calculations in order to assess exactly
how correlated U is with respect to other actinide metals.  From a
theoretical point of view, the correlations of U, while somewhat
strong, may yet be weak enough to be tractable by modern many-body
techniques such as dynamical mean-field theory (DMFT)~\cite{DMFT}
and may be far easier to understand than more strongly correlated
materials like Pu. However, the first step in this process is to
establish high-quality photoemission spectra for very good single
crystals and compare these results with band-structure calculations
in order to provide a reliable baseline for whatever correlations
are present. This paper provides preliminary results in this
direction.

Uranium, the heaviest natural element, exists in three allotropes
and has a complex phonon spectrum~\cite{manley1} and electronic
structure. Unusual properties of uranium also include anisotropic
thermal expansion~\cite{Barrett, Lloyd,Lawson}, the occurrence of
three charge-density wave (CDW) transitions~\cite{Berlincourt, Fisher170, Nelson1}
below 43~K, and strongly temperature-dependent elastic
moduli~\cite{4, McSkimin}. Aside from the low-temperature CDW transitions, the
ground-state structure for uranium is orthorhombic ($\alpha$-U).
Upon heating, $\alpha$-U transforms into a tetragonal structure
(\textit{T}$_\beta$ = 935~K) and finally crystallizes to a
body-centered cubic phase (\textit{T}$_\gamma$ = 1045~K) prior to
melting at 1406~K, all at ambient pressure~\cite{Lawson,5}. Many of
the unusual properties found in uranium, as with the other light
actinides (Th-Pu)~\cite{6}, are thought to be related  to the
delocalization of the partially filled U5f electronic states and
their hybridization with the U6d-7s electronic states~\cite{7}. The
U5f electrons participating in bonding have been shown in uranium
intermetallics to exhibit magnetism and superconductivity~\cite{8}
and show similar bonding behavior to the $d$ electrons in lanthanide
and transition metals~\cite{9}.

Several photoemission experiments have been carried out on uranium~\cite{Gouder295,Gouder341_382Moldtsov,arko,Schneider22}.  Unfortunately, these
experimental studies often suffer from poor spectral resolution
caused by either oxygen contamination or the use of samples created
by metal deposition upon a substrate.  Thin-film deposition studies,
although valuable, might not be truly representative of a bulk
material.  The electronic structure of thin films is influenced by
the chemical interaction between the overlayer and the
substrate. Using large U single crystals and a thorough
sputter-anneal regimen, we have overcome these difficulties.

In this paper we present valence-band photoemission spectra at HeI
and HeII energy excitations for a very high-quality single crystal
of U at 173K and compare these with the results of first principles
calculation of the electronic structure using the generalized
gradient approximation approach (GGA) in the full-potential
linearized-augmented-plane-wave (FLAPW) method, which includes local
6p orbitals to accommodate the low-lying 6p semicore
states~\cite{Blaha}. Using XPS we explore U6p states and note a
splitting in the 6p$_{3/2}$ manifold indicative of a core-valence
band separation due to hybridization. The normal-incidence U(001)
photoemission spectroscopy and LEED results confirm that our U
single-crystal surface shows long-range order and is representative
of the bulk.

\begin{figure}[b]
 \includegraphics[width=\columnwidth]{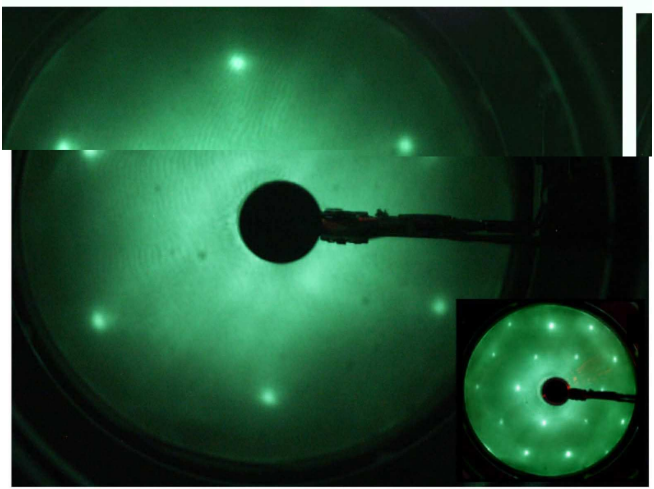}
   \caption{\label{fig:1} (Color online)
   Single crystal LEED pattern of a clean first order $1\times 1$ U
   (001) surface at normal incidence and an electron-beam energy of
   $\approx$~50~eV at room temperature. Inset shows higher order LEED
   pattern at $\approx$~150~eV.
}
\end{figure}

\section{Experiment}

Single crystals of $\alpha$-U were grown at Argonne National
Laboratory by electro-refinement in a molten (LiCl-KCl) eutectic
electrolyte containing 3 wt.~\% UCl$_{3}$ at 773~K~\cite{11}. This
procedure grows the crystals directly in the $\alpha$-phase and
avoids the formation of high-temperature structures. The crystals
(as large as 10$\times$10$\times$1~mm$^3$) collect on a stainless
steel cathode as dendrites or thin parallelogram-shaped platelets.
To remove any residual salt, the U crystals were cleaned with water
and electropolished in H$_3$PO$_4$ prior to the experiment. Chemical
analysis reports 40 (atomic)~ppm C and 167 (atomic)~ppm Si as the
only detectable impurities.

Unlike previous U samples\cite{Lander43}, these crystals are easily
bent, and small cross-section pieces can be deformed by rotating a
necked region by hand through several turns without work hardening
or weakening. Because these crystals have no grain boundaries and
few impurities, we suggest that this unique ductility is the result
of a large number of potential twin planes of the orthorhombic
structure, and the ability of the twin to move over
millimeters~\cite{Cahn49DanielHuddant}. Characterization by X-ray
diffraction Laue patterns found no detectable structural
imperfections and show that the c--axis is perpendicular to the
platelet surface.

In previous resistivity studies~\cite{Lashley, Schmiedeshoff}
single crystals from the same source, the crystals were found to
have a residual resistivity ratio of up to 315, eight times
higher than previously reported values~\cite{Berlincourt, Brodsky}.
We take the above as evidence that these are the highest quality
single $\alpha$-U crystals yet produced, and that they possess
extremely low impurity concentration and minimal micro-structural
defects.

Ultraviolet photoelectron spectra were recorded with a resolution of 28.5~meV
using a Perkin-Elmer/Physical Electronics Model 5600 ESCA system
equipped with a monochromated Al~\textit{K}$\alpha$ (1486.6~eV),
a SPECS UVS 300 ultraviolet lamp (HeI, $h\nu$ = 21.21~eV, HeII,
$h\nu$ = 40.81~eV), and a spherical capacitor analyzer. The vacuum
chamber, which had a base pressure of $1.3 \times 10^{-8}$ Pa, was
equipped with a variable temperature sample stage of the range
150--1273~K.  Our crystal surface was aligned perpendicular to the
analyzer and set at an acceptance angle of $\pm$2 degrees in order
to produce greatest sensitivity.  Surface preparation for both
spectroscopic and LEED measurements consisted of repeated cycles of
Ar ion sputtering and annealing at 873~K.  After preparation, the
oxygen (O1s) and carbon (C1s) signals in the XPS spectra, major
contaminant indicators on metallic actinide surfaces, were below the
detection limit ($<$~1~at.~\%).

\section{U(001) LEED measurements}

In an effort to determine sample surface quality, we performed
Low-Energy Electron Diffraction (LEED) measurements on our samples
using an Omicron Spectraleed analyzer
with the electron beam at normal incidence.  We show in
Fig.~\ref{fig:1} the first reported LEED of long-range order in a
U(001) single crystal surface structure at room temperature with an
electron energy of 50~eV.  Higher order reciprocal space LEED
patterns, up to third order, were clearly visible at greater
energies, see inset Fig.~\ref{fig:1}.  We find no evidence of
surface reconstruction, and analysis on the bulk termination (1x1)
LEED pattern confirms it is consistent ($<~$2~\% difference) with the
diffraction pattern calculated for an orthorhombic U(001)
crystallographic structure (\textit{a} = 2.8537~\r{A}, \textit{b}  =
5.8695~\r{A}, \textit{c}  = 4.9548~\r{A}) at room
temperature~\cite{Barrett}.

The quality and character of the sample surface is of critical
importance for conducting electron-structure measurements. Due to
the strong chemical reactivity of uranium, Ar sputtering was
utilized to prepare a clean surface, and confirmed by XPS, prior to
each measurement~\cite{Cardona}. We found that the Ar ion sputter
damage from cleaning the crystal surface was removed by annealing at
873K for a few minutes and then reducing the temperature to 673~K.
After this temperature sequence, the surface re-ordered, and a
distinct U(001) diffraction pattern appears.

\section{UPS: valence band spectra and DOS calculation}

\begin{figure}[t!]
\includegraphics[width=\columnwidth]{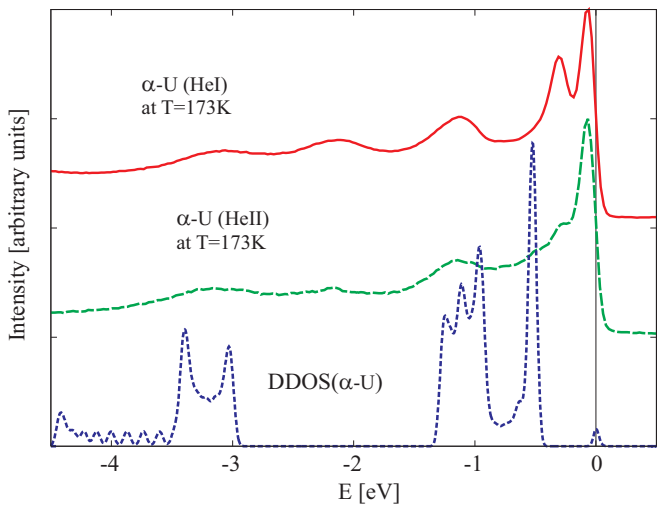}
   \caption{\label{fig:3m} (Color online)
   Intensity/DDOS as a function of binding energy (eV)
   for UPS (HeI, $h\nu$ = 21.2~eV, and HeII, $h\nu$ = 40.8~eV)
    valence band data on single crystal U(001) at T = 173 K. }
\end{figure}

In the past, UPS measurements for most light actinides supplied only
a familiar triangle shaped peak close to the Fermi
edge~\cite{Gouder295,Gouder341_382Moldtsov,VealGouder271Laubschat,McLean}.
Given our sample quality, alignment, and enhanced resolution we are
able to discern more structure in the valence band.  An expanded
view of our UPS valence band data for $\alpha$-U at \textit{T} =
173~K is depicted in Fig.~\ref{fig:3m}.
The background from inelastic scattering of secondary electrons in the HeI spectra was removed by subtraction of an exponential function from below $E_F$ to the peak of the background.
Comparing the HeI and HeII
spectra at 173 K, we note that almost all spectral features (peaks)
line up in both experimental spectra. The difference in relative
intensities between the two spectra has to do with the different
cross sections between $d$ and $f$ states, different escape depths
of the excited electrons, and other factors, which we will discuss
below.

In Fig.~\ref{fig:3m} we also present results from first-principles
GGA~WIEN2K\cite{Blaha} electronic band-structure calculations.  The theory
is based upon the simple notion that only ${\mathbf{k}}_{\|}$ is
conserved, and hence for normal photoemission all electronic states
along the direction $\Gamma$ to Z are present (${\mathbf{k}}_{\bot}$
is not conserved). Hence, the theoretical curve is a directional
density of states (DDOS) as a function of energy $E$, which is
calculated from
\begin{equation} \label{eq:PDOS}
  \mathrm{DDOS}(E) = \sum_{{\mathbf{k}}=\Gamma}^Z
  \sum_\lambda  \delta(E-E_{{\mathbf{k}},\lambda}) f(E-E_F)~,
\end{equation}
where $E_{{\mathbf{k}},\lambda}$ is the energy eigenvalue for
${\mathbf{k}}$, band-index $\lambda$, $f(E-E_F)$ is the Fermi
function for electron occupancy, and $E_F$ is the Fermi energy.  In
this formula we have used a Dirac delta function for the
contribution to the DDOS for each band state.  Since we have only
summed over 21 ${\mathbf{k}}$-points between $\Gamma$ to Z, it was
necessary to broaden the delta function into a finite Gaussian in
order to draw a smooth curve. We used a full-width at half max of
28.5 meV for the Gaussian (the instrumental resolution of the
experiment). The wiggles between -3.5 to -4.5 eV show the coarseness
of our ${\mathbf{k}}$-point grid versus Gaussian width. If we wished
to smooth out this part of the DDOS, we could either increase the
width of the Gaussian or the number of ${\mathbf{k}}$-points.

\begin{figure}[b!]
\includegraphics[width=\columnwidth]{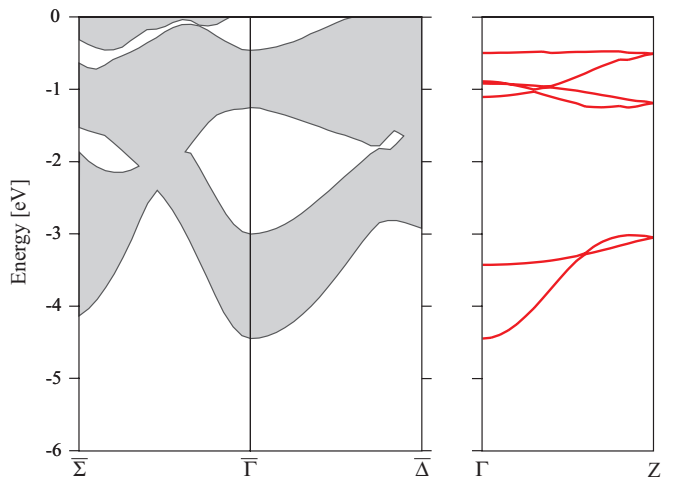}
   \caption{\label{fig:3n} (Color online)
   GGA band-structure calculations for $\alpha$-U.  The shaded region of
   the left hand part of the figure indicates the range of values where
   energy bands exist when projected on the $\Gamma$ to $\Delta$
   and $\Gamma$ to $\Sigma$ directions.  The white areas indicate
   possible regions where surface states might exist.
   Note that at normal incidence in the U(001) plane,
   surface states are possible in the region
   between $E_F$ and about -0.5 eV, and from about 1.5 to 3 eV
   below $E_F$.  These regions are also free of energy states
   for the energy bands along the $\Gamma$ to Z direction, which are
   shown in the right-hand side panel of the figure.
}
\end{figure}

Besides spectrometer resolution effects, each eigenvalue
$\delta$-function should actually have a width representative of the
lifetime of the hole state (due to radiative and Auger decay).  The
lifetime, which is of the order of $\hbar$/width, should be
increasingly shorter for higher binding energy; a simple
free-electron argument would give a Gaussian width for each state of
energy E proportional to $(E-E_F)^2$. Since it is very difficult to
calculate hole lifetimes from first principles, we have not included
this effect in Fig.~\ref{fig:3m}. The net effect of including
lifetimes would be to progressively smear out all the theoretical
features as one moved to higher binding energy (below the Fermi
energy).  This effect is clearly seen in the experimental spectra.
Also note that the peaks in the DDOS correspond to flat regions of
the energy bands (small dispersion) along $\Gamma$ to Z (cf. the
band states on the right-hand side of Fig.~\ref{fig:3n}).

\begin{figure}[b!]
\includegraphics[width=\columnwidth]{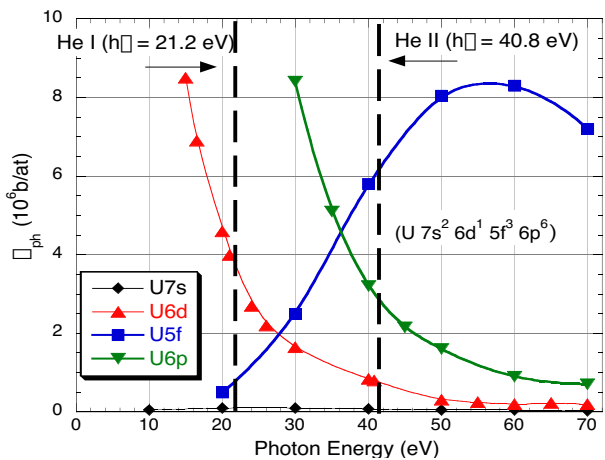}
   \caption{\label{fig:4} (Color online)
Atomic photoionization cross section~\cite{Yeh} vs. photon energy
for U 7s$^2$ 6d$^1$ 5f$^3$ 6p$^6$ energy bands. }
\end{figure}

A comparison between the band-structure results and the experimental
spectra in Fig.~\ref{fig:3m} shows favorable agreement for the peaks
near -1.2 eV and -3.2 eV. The region between 0 and 1 eV below $E_F$
has mainly $f$-electron character. We expect that these types of
states (especially near $E_F$) should show the largest effects due
to electron-electron correlations that go beyond those included in
LDA (or GGA) band-structure calculations. Therefore, we argue that
the theoretical peak near -0.6 eV likely corresponds to the
experimental peak near -0.3 eV, and that the shift is likely a
quasi-particle effect due to these additional electronic
correlations. (This effect will be the subject of a forthcoming
study in the DMFT framework.) The two remaining peaks (at -0.1 and
-2.2 eV) appear at gaps in the conventional band structure (see the
left-hand side of Fig.~\ref{fig:3n}), and therefore are likely
surface states of mainly $f$ and $d$ character, respectively.

To understand the relative intensities of the various peaks between
the HeI and HeII spectra is somewhat complicated. According to band
theory, the peaks in the spectra come from high projected densities
of states, which arise from flat regions of the bands (the flatter
the bands the sharper the peak). In addition, because of cross
section effects (see below), any $d$-electron feature will be
enhanced in the HeI spectra and any $f$-electron feature will be
enhanced in the HeII spectra. In addition, we estimate that the
electronic mean-free path is probably close to its minimum value for
the HeII spectra, and hence any surface state will be enhanced
relative to bulk states for the HeII spectra. Since both spectra are
normalized to the maximum intensity in the Fermi-energy region and
not absolute values, only relative peak heights within each spectra
have meaning, and we cannot compare absolute values between the two
spectra.  In addition, as discussed above, peaks at higher binding
energy are due to electronic energy states that have much shorter
lifetimes (due to radiative decay and Auger mechanisms), which
broaden these states and lower the intensity of the peaks.

Given the large number of factors in determining the relative height
of each peak, only qualitative statements can be made: The peaks
below -1.0 eV show up much more prominently in the HeI spectra
relative to the HeII spectra, because the $d$ electron photoemission is
enhanced.  This can be seen from the atomic photoionization
cross-sections~\cite{Yeh}, which are shown in Fig.~\ref{fig:4}. Note
that the HeI data strongly emphasize the $d$ electrons and the HeII
data the $f$ electrons.

The -0.1 eV peak in the HeII spectra is very enhanced.  Because this
is likely a $f$-character surface state, there are several possible
contributing factors to its strength: First, an $f$-electron surface
state will have reduced hybridization and a high one-electron
density of states. Surface atoms have a smaller number of near
neighbors, which causes a higher local DOS for these atoms.
Correlation effects are likely to increase this DOS.  Also, the
$f$-electron cross sections are very strong for the HeII spectra.
Secondly, according to estimates for escape depth as a function of
excited electron energy based on the universal curve\cite{UniversalCurveRef}, we believe that the inelastic mean free path of the
excited electrons for the HeII spectra should be near an absolute
minimum and should thus be smaller than for the HeI spectra,
which should enhance surface state features in the HeII spectrum.

In contrast, the peak at -2.2 eV is likely a surface state with $d$
character. Due to the interplay of the photoionization
cross-sections, the peak is emphasized in the HeI spectrum and
suppressed in the HeII, the latter effect being enhanced by the fact
that the spectra are normalized at the maximum intensity.

From our band-structure results for the total DOS at the Fermi
energy, we can estimate the effective mass enhancement $\lambda$ by
comparing to specific-heat measurements. We find $\lambda =
(\gamma_{\mathrm {exp}} /\gamma_{\mathrm {cal}}) - 1$ to be 0.55,
consistent with a previous calculation by Skriver et
al.~\cite{Skriver32}. Our DOS calculation is similar to those
previously computed for $\alpha$-U by Wills and
Eriksson~\cite{Wills}, and P\'enicaud~\cite{Penicaud}. There are two
general contributions to the effective-mass enhancement:
electron-phonon and electron-electron.  A many-body theory that is
beyond the scope of this paper would be required to sort out the
relative contributions.

\section{X-ray photoemission spectroscopy and U6p electron bands}

\begin{figure}[t!]
 \includegraphics[width=\columnwidth]{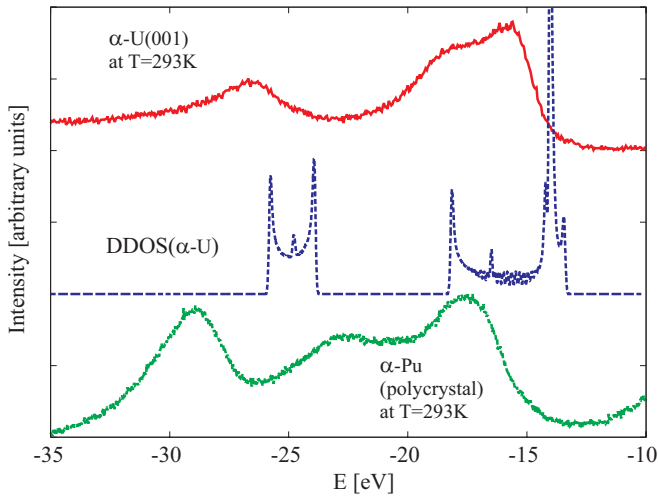}
   \caption{\label{fig:5} (Color online)
Comparison of XPS-HRES (Al K-$\alpha$, 1486.6~eV) spectra for
U6p$_{1/2}$ - 6p$_{3/2}$, U6p DDOS calculation, and XPS-HRES (Al
K-$\alpha$, 1486.6~eV) spectra for Pu6p$_{1/2}$ -
6p$_{3/2}$~\cite{Tobin}. For the DDOS calculation we used the XPS
resolution of 50~meV.}
\end{figure}

Figure~\ref{fig:5} shows a comparison of U6p$_{1/2}$--6p$_{3/2}$ XPS
spectra at room temperature, a theoretical $\alpha$-U  DOS
(\textit{T} = 0~K) calculation, and XPS data for
Pu6p$_{1/2}$--6p$_{3/2}$ previously reported by Tobin \emph{et
al.}~\cite{Tobin}. We note that the spin-orbit splitting ($\approx$
9.5~eV) between U 6p$_{1/2}$--6p$_{3/2}$ data corresponds well with
the DOS calculation. The broadness of the peak at 27~eV is due to a
combination of thermal broadening and the considerable
quasi-particle lifetime effects for states so far below E$_F$. Much
theoretical work~\cite{Singh,kunes,Nordstrom} has been done evaluating the 6p
states in the light actinides to model accurately spin-orbit
coupling.  In comparison with data of other actinide metals, a spin-orbit
splitting of similar order has been observed in Th by Fuggle
\emph{et al.}~\cite{Fuggle4}, as well as in the theoretical
calculation by Kune\v{s} \emph{et al.}~\cite{kunes}, and in the Pu
data~\cite{Tobin}.

We note that a shoulder emerges on the left side of the U6p$_{3/2}$
data peak ($\approx$18.6~eV) in Fig.~\ref{fig:5}, which show the
hybridization features of this band-like shallow core state. This new observation
may be directly attributed to the purity of our $\alpha$-U single
crystal with minimum strain and low-impurity concentrations.
Previous XPS experiments on uranium thin film~\cite{Fuggle4} and
polycrystal~\cite{McLean} samples fail to indicate this shoulder in
the U6p$_{3/2}$ peak intensity at 18.6~eV as shown in our single
crystal data. Comparison with the DOS calculation indicates a clear
splitting of the U6p$_{3/2}$ over a similar energy range as seen in
the U data.  Normally, one expects the U6p electrons more than 15 eV away
from the Fermi edge (E$_{F}$) to exhibit exclusively  core-like behaviors.  However,
 the splitting of the U data peak and confirmation via calculation lead us
 to speculate certain electrons may hybridize.  Hybridization between 6p and 6d electrons
 is allowed via j-j coupling and is supported by applicable symmetry rules.\cite{Nunes}  Studies involving
 density functional theory (DFT)~\cite{Ermler, Hay} argue persuasively to include U6p electrons in the valence
 band, and the clear overlap of radial distribution functions for light actinides may increase the
 likelihood of such hybridization~\cite{Batista}.  Thus our data may constitute the first experimental evidence
 for such 6p and 6d hybridization in the condensed phase of actinide metals.

Although no experimental evidence of the Th6p$_{3/2}$ electron
splitting is currently available, Kune\v{s} \emph{et
al.}~\cite{kunes} have calculated this using a similar GGA FLAPW
approach. This calculation unambiguously shows Th6p$_{3/2}$ peak
splitting over a 2~eV energy interval.  Evidence of a similar
6p$_{3/2}$ splitting is visible in the Pu data~\cite{Tobin}.

In order to exclude the possibility that the observed splitting of
the U6p$_{1/2}$--6p$_{3/2}$ peaks is the result of surface
reconstruction, relaxation, or contamination effect, an oxidized
U(001) sample surface was cleaned in stages via Ar sputtering and
analyzed with HRES--XPS. As the O1s (531~eV) peak was eliminated,
the oxide (U$_x$O$_y$) peaks associated with the valence band (-29
and -24.5~eV) simultaneously dissipated. As sputtering continued the
6p$_{1/2}$ and 6p$_{3/2}$ peaks emerge at -26.8 and -17.0~eV. These remain when the sample is annealed up to 873~K in order to reorder the surface atoms, and surface impurity is below detectability.  From
this result we conclude that the U6p$_{1/2}$ - 6p$_{3/2}$
photoemission measurements are representative of the bulk and
preclude any anomalous surface reconstruction effect. Subsequent
experiments on other high-quality polycrystal U indicate that the
shoulder on the U6p$_{3/2}$ data peak and the
U6p$_{1/2}$--6p$_{3/2}$ spin-orbit splitting remain at 9.5~eV up to
1100~K.

\section{Conclusions}

In this paper we present the first U(001) LEED pattern corresponding
to long-range order in a uranium single-crystal surface. We report
favorable agreement between first-principles GGA band structure
calculations and the valence band UPS data. We also identify peaks
which likely correspond to surface states present in the gaps of the
conventional band structure, at normal incidence on the U(001)
plane. We note that at the higher binding energies $\approx$
13--30~eV using XPS, the GGA band structure correctly predicts the
behavior of the U6p$_{1/2}$--6p$_{3/2}$ core states, showing both
the spin-orbit splitting (9.5~eV) and hybridization effects.

To our knowledge, with the exception of the recent EXAFS studies~\cite{Nelson1}, single crystals of this quality have not been
previously utilized for surface spectroscopy, and because of their
purity, the photoemission results show many more features than
previous experiments, providing new insight into the
electronic-structure of $\alpha$-U. Angle-resolved photoemission
spectroscopy (ARPES) experiments are currently underway to map the
band structure of these $\alpha$-U single crystals. ARPES
measurements are required to study the bands' dispersion, and will
allow for a detailed comparison with first-principle GGA band
structure calculations. To minimize the spectrum contamination due
to surface-state effects, the ongoing ARPES experiments are being
performed at HeI photon energy. Finally, these ARPES measurements
will also help study the character of the features which were
tentatively identified as surface states in the present study.


\begin{acknowledgments}
This work was supported in part by the LDRD program at Los Alamos
National Laboratory.  We gratefully acknowledge the contribution of
the U(001) samples from: H. F. McFarlane, K. M. Goff, F. S.
Felicione, C. C. Dwight, D. B. Barber, C. C. McPheeters, E. C. Gay,
E. J. Karell and J. P. Ackerman at Argonne National Laboratory.
B.M. acknowledges help from M. D. Jones and I.~Schnell in getting acquainted with the
WIEN2K package.
\end{acknowledgments}

\vfill


%
%
%
%
%

\end{document}